# A General and Yet Efficient Scheme for Sub-Nyquist Radar Processing

Shengyao Chen, Feng Xi, Zhong Liu[*]

Department of Electronic Engineering, Nanjing University of Science and Technology, China

**Abstract**—We study the target parameter estimation for sub-Nyquist pulse-Doppler radar. Several past works have addressed this problem but either have low estimation accuracy for off-grid targets, take large computation load, or lack versatility for analog-to-information conversion (AIC) systems. To overcome these difficulties, we present a general and efficient estimation scheme. The scheme first formulates a general model in the sense that it is applicable to all AICs regardless of whether the targets are on or off the grids. The estimation of Doppler shifts and delays is performed sequentially, in which the Doppler estimation is formulated into a spatial spectrum estimation problem and the delay estimation is decomposed into a series of compressive parameter estimation problems with each corresponding to an estimated Doppler shift. By the sequential and decomposed processing, the computational complexity is substantially reduced, and by the parametric estimation techniques, the high accurate estimation is obtained. Theoretical analyses and numerical experiments show the effectiveness and the correctness of the proposed scheme.

## I. Introduction

We consider target parameter estimation in a sub-Nyquist pulse-Doppler radar via analog-to-information conversion (AIC) systems. The problem has received extensive attention in recent years and several methods have been reported [1-6]. Among these reported methods, some estimate the target parameters simultaneously by sparse reconstruction [1-3], some perform sequential estimation of the delays and Doppler shifts [4-6], some assume that the targets are located on the predefined delay-Doppler grids [1, 2, 4, 5], and some are developed on specific AIC systems [4-6]. These methods either have low estimation accuracy for off-grid targets, take large computation load, or lack versatility for AIC systems (See [7] for a review). In general, the sequential methods have computational advantages and are more suitable to practical radar applications.

Recently, we developed a general sequential delay-Doppler (GeSeDD) estimation scheme in [7]. GeSeDD is versatile in suiting for any AIC system and has the advantages of high estimation accuracy, high resolution and moderate-size computational complexity. This paper continues our study on the development of general sequential estimation method with the aim to reduce the computational overhead. Instead of grouping the targets into classes with each corresponding to a distinct time delay [7], we classify the targets into groups with each corresponding to a distinct Doppler shift. For the universal applicability, we formulate a compressive measurement model regardless of AIC systems and on-grid and/or off-grid targets. By the new model, we find that the Doppler shifts can be estimated independently of the time delays by parametric direction-of-arrival (DOA) estimation techniques [8]. With the estimated Doppler shifts, we establish a tactics that further decomposes the estimation of time delays into a series of compressive parameter estimation problems with each corresponding to a distinct Doppler shift. The compressive parameter estimation techniques [3, 9-11] can be taken to estimate the time delays. By the sequential and decomposed processing, combined with parametric estimation techniques, the novel estimation scheme has high computational efficiency and high estimation accuracy. Theoretical analyses and simulation results confirm our assertions.

---

[*] Corresponding author. Email address: eezliu@njust.edu.cn

The rest of this paper is organized as follows. The problem formulation is defined in Section II. The estimation scheme of Doppler shifts and delays is introduced in Section III. Section IV presents the analyses on computational complexity. The experiments are simulated in Section V. Conclusion is given in Section VI.

## II. Problem Formulation

The data model we will develop is similar to that in [7], except that we classify the targets into groups with each corresponding to a distinct Doppler shift. Consider a radar scene consisting of $K$ non-fluctuating moving point targets. Assume that there are $K_v$ ($K_v < L$) targets having distinct Doppler shifts and each Doppler shift $v_i$ is associated with $K_{v,i}$ different delays $\tau_{ij}$ ($j=1,\cdots,K_{v,i}$). For a co-located pulse-Doppler radar transmitting $L$ pulses in a coherent processing interval (CPI), the complex baseband echo corresponding to the $l$-th pulse ($0 \le l \le L-1$) is given by

$$\tilde{r}^l(t) = \sum_{i=1}^{K_v} e^{j2\pi v_i t} \sum_{j=1}^{K_{v,i}} \alpha_{ij}\tilde{g}(t-lT-\tau_{ij}) \approx \sum_{i=1}^{K_v} e^{j2\pi v_i lT} \sum_{j=1}^{K_{v,i}} \alpha_{ij}\tilde{g}(t-lT-\tau_{ij}), \quad t \in [lT,(l+1)T) \tag{1}$$

where $T$ is the pulse repetition interval (PRI), $\tilde{g}(t)$ is the complex envelop of transmitting pulses with bandwidth $B$ and pulse width $T_p$ ($T_p < T$) and $\alpha_{ij} \in \mathbb{C}$ is the complex gain of the target with the delay-Doppler pair $(\tau_{ij}, v_i)$. The approximation follows from the stop-and-hop assumption [12]. For unambiguous estimation, the delays and Doppler shifts satisfy $\tau_{ij} \in [0, T-T_p)$ and $v_i \in (-1/2T, 1/2T)$.

Under parametric waveform dictionary [13] with the delay $\tau_{ij}$, $\{\tilde{\psi}(\tau_{ij},t)|\tilde{\psi}(\tau_{ij},t)=\tilde{g}(t-\tau_{ij}), i=1,\cdots,K_v, j=1,\cdots,K_{v,i}\}$, the echo signal $\tilde{r}^l(t)$ can be expressed as

$$\tilde{r}^l(t) = \sum_{i=1}^{K_v} e^{j2\pi v_i lT} \sum_{j=1}^{K_{v,i}} \alpha_{ij} \tilde{\psi}(\tau_{ij}, t-lT), \qquad t-lT \in [0,T) \tag{2}$$

In the presence of noise, the received baseband signal is given by

$$\tilde{r}_1^l(t) = \sum_{i=1}^{K_v} e^{j2\pi v_i lT} \sum_{j=1}^{K_{v,i}} \alpha_{ij} \tilde{\psi}(\tau_{ij}, t-lT) + \tilde{n}^l(t), \qquad t-lT \in [0,T) \tag{3}$$

where $\tilde{n}^l(t)$ is assumed to be a lowpass complex Gaussian process with power spectral density $N_0$ and bandwidth $B$, and all of the $\tilde{n}^l(t)$ ($1 \leq l \leq L$) are independently and identically distributed (i.i.d.).

For sub-Nyquist radars, the received signal $\tilde{r}_1^l(t)$ is sampled by an AIC system to attain compressive measurements [14]. Denote $\mathbf{r}^l = [\tilde{r}^l(lT), \tilde{r}^l(lT+T_{nyq}), \cdots, \tilde{r}^l(lT+(N-1)T_{nyq})]^T \in \mathbb{C}^{N \times 1}$ and $\mathbf{n}^l = [\tilde{n}^l(lT), \tilde{n}^l(lT+T_{nyq}), \cdots, \tilde{n}^l(lT+(N-1)T_{nyq})]^T \in \mathbb{C}^{N \times 1}$ as the Nyquist-rate sampling vectors of $\tilde{r}^l(t)$ and $\tilde{n}^l(t)$ in a PRI $t \in [lT, (l+1)T)$, respectively, where the superscript $(\cdot)^T$ represents transposition. The compressive measurement vector $\mathbf{s}_{cs}^l \in \mathbb{C}^{M \times 1}$ of the signal $\tilde{r}_1^l(t)$ obtained from any AIC system can be represented as

$$\mathbf{s}_{cs}^l = \mathbf{M}(\mathbf{r}^l + \mathbf{n}^l) = \mathbf{M}\mathbf{r}^l + \mathbf{n}_{cs}^l \tag{4}$$

where $\mathbf{M} \in \mathbb{C}^{M \times N}$ ($M < N$) is the measurement matrix and $\mathbf{n}_{cs}^l = \mathbf{M}\mathbf{n}^l = [\tilde{n}_{cs}^l[1], \tilde{n}_{cs}^l[2], \cdots, \tilde{n}_{cs}^l[M]]^T \in \mathbb{C}^{M \times 1}$ is the compressive measurement vector of the noise $\tilde{n}^l(t)$. The $\tilde{n}_{cs}^l[m]$ ($1 \leq m \leq M$) is an i.i.d. complex Gaussian process with zero-mean and variance $NN_0B/M$ [15]. In the formulation of (4), we are not confined to any specific AIC system. Then the matrix $\mathbf{M}$ depends on the assumed AIC system. In this sense, (4) is a general compressive measurement model for any sub-Nyquist radar.

Let $\mathbf{\psi}(\tau_{ij}) = [\tilde{\psi}(\tau_{ij}, 0), \tilde{\psi}(\tau_{ij}, T_{nyq}), \cdots, \tilde{\psi}(\tau_{ij}, (N-1)T_{nyq})]^T \in \mathbb{C}^{N \times 1}$ be the Nyquist-rate sampling vector of the parametric atom $\tilde{\psi}(\tau_{ij}, t)$. The compressive vector $\mathbf{s}_{cs}^l$ can be explicitly described by

$$\mathbf{s}_{cs}^l = \sum_{i=1}^{K_v} e^{j2\pi v_i lT} \sum_{j=1}^{K_{v,i}} \alpha_{ij} \mathbf{M}\mathbf{\psi}(\tau_{ij}) + \mathbf{n}_{cs}^l = \sum_{i=1}^{K_v} e^{j2\pi v_i lT} \mathbf{M}\mathbf{\psi}_i + \mathbf{n}_{cs}^l \tag{5}$$

where $\mathbf{\psi}_i = \sum_{j=1}^{K_{v,i}} \alpha_{ij} \mathbf{\psi}(\tau_{ij}) \in \mathbb{C}^{N \times 1}$. Define $\mathbf{\Psi} = [\mathbf{\psi}_1, \mathbf{\psi}_2, \cdots, \mathbf{\psi}_{K_v}] \in \mathbb{C}^{N \times K_v}$ and $\mathbf{\theta}^l = [e^{j2\pi v_1 lT}, e^{j2\pi v_2 lT}, \cdots, e^{j2\pi v_{K_v} lT}]^T \in \mathbb{C}^{K_v \times 1}$. We can rewrite (5) as

$$\mathbf{s}_{cs}^l = \mathbf{M}\mathbf{\Psi}\mathbf{\theta}^l + \mathbf{n}_{cs}^l \tag{6}$$

Concatenating all $L$ compressive measurement vectors $\mathbf{s}_{cs}^l$ and $\mathbf{n}_{cs}^l$ as $\mathbf{S} = \left[\mathbf{s}_{cs}^0, \mathbf{s}_{cs}^1, \cdots, \mathbf{s}_{cs}^{L-1}\right] \in \mathbb{C}^{M \times L}$ and $\mathbf{N} = \left[\mathbf{n}_{cs}^0, \mathbf{n}_{cs}^1, \cdots, \mathbf{n}_{cs}^{L-1}\right] \in \mathbb{C}^{M \times L}$, respectively, we have

$$\mathbf{S} = \mathbf{M}\mathbf{\Psi}\mathbf{\Theta} + \mathbf{N} \tag{7}$$

with $\mathbf{\Theta} = \left[\mathbf{\theta}^0, \mathbf{\theta}^1, \cdots, \mathbf{\theta}^{L-1}\right] \in \mathbb{C}^{K_v \times L}$.

Our goal is to accurately and efficiently estimate the target parameters $\tau_{ij}$, $\alpha_{ij}$ and $v_i$ ( $i = 1, 2, \cdots, K_v$, $j = 1, 2, \cdots, K_{v,i}$ ) from the compressive measurement data matrix $\mathbf{S}$. As discussed in [7], estimating these parameters can be performed by solving a high-dimensional nonlinear optimization, which will take large computational load.

### III. Estimation of target parameters

In the proposed scheme, we first estimate the Doppler shifts and then the delays and gains.

*A. Estimation of the Doppler shifts*

The estimation of Doppler shifts is performed independently of the delays and gains. To see this, we take the transpose operation on both sides of (7) and have

$$\mathbf{S}^{\mathrm{T}} = \mathbf{\Theta}^{\mathrm{T}}\mathbf{\Psi}^{\mathrm{T}}\mathbf{M}^{\mathrm{T}} + \mathbf{N}^{\mathrm{T}} \tag{8}$$

where $\mathbf{\Theta}^{\mathrm{T}} = \left[\mathbf{a}(v_1), \mathbf{a}(v_2), \cdots, \mathbf{a}(v_{K_v})\right] \in \mathbb{C}^{L \times K_v}$ and $\mathbf{a}(v_i) = \left[1, e^{j2\pi v_i T}, \cdots, e^{j2\pi v_i (L-1)T}\right]^{\mathrm{T}} \in \mathbb{C}^{L \times 1}$. If we take $\mathbf{\Theta}^{\mathrm{T}}$ as the steering matrix and $\mathbf{\Psi}^{\mathrm{T}}\mathbf{M}^{\mathrm{T}}$ as the source matrix consisting of $K_v$ sources, we can see that (8) is exactly the same as the model of classical DOA estimation problem for an $L$-element uniform linear array [8]. The $i$-th row of $\mathbf{\Psi}^{\mathrm{T}}\mathbf{M}^{\mathrm{T}}$ represents the arriving source with spatial frequency $f_i = v_i T \in (-1/2, 1/2)$, and the number of snapshots equals to that of compressive samples $M$ in one PRI. The target parameters, gains and delays, are contained in $\mathbf{\Psi}$ which does not affect the independent estimation of the Doppler shifts. Then the Doppler shifts can be estimated by the parametric DOA estimation methods, such as MUSIC [16] and ESPRIT [17].

In practical radar scenarios, there may be several targets with the same delays but having different Doppler shifts. In such case, the source signals resulting from the radar echoes become coherent and the resulting matrix $\mathbf{\Psi}^{\mathrm{T}}\mathbf{M}^{\mathrm{T}}$ is not of full-row rank. If the case occurs, we can take the coherent DOA techniques [18] to estimate the Doppler shifts.

The parametric DOA estimation idea has been taken in our previous work [7] which formulates the estimation of the time delays as a DOA estimation problem. However, the proposed method is superior to the GeSeDD method when the coherent sources appear in the corresponding DOA estimation problem. In GeSeDD, the estimation of time delays is targeted as a beamspace DOA estimation problem, where classical coherent DOA estimation techniques cannot be exploited directly [8] and advanced techniques, such as interpolated array technique [19], should be utilized. Then the computational load increases and the estimation accuracy is also affected. In contrast, the proposed method can figure out the coherent sources by classical techniques, since the DOA estimation problem in (8) is of element-space.

When the Doppler shifts are all on the Nyquist grids, the proposed method reduces to the compressive sampling pulse-Doppler (CoSaPD) method [4] if we take the discrete Fourier transform (DFT) on (8) to estimate the Doppler shifts. In this case, the coherent sources have no effect on the estimation of Doppler shifts.

*B. Estimation of the delays and gains*

With the estimated Doppler shifts, a straightforward way to estimate the delays $\tau_{ij}$ and gains $\alpha_{ij}$ ($i=1,2,\cdots,K_v$, $j=1,2,\cdots,K_{v,i}$) is to solve the following optimization problem

$$\arg\min_{\alpha_{ij},\tau_{ij}} \|\mathbf{S} - \mathbf{M\Psi\Theta}\|_F^2 \tag{9}$$

By expressing $\mathbf{M\Psi\Theta}$ as $\mathbf{M\Psi\Theta} = \sum_{i=1}^{K_v} \mathbf{M}\boldsymbol{\psi}_i \mathbf{a}(v_i)^T$ and column-wise vectorizing the equation (7), we can translate the problem (9) into

$$\arg\min_{\alpha_{ij},\tau_{ij}} \left\| \text{vec}(\mathbf{S}) - \sum_{i=1}^{K_v} \mathbf{a}(v_i) \otimes \mathbf{M} \sum_{j=1}^{K_{v,i}} \alpha_{ij}\boldsymbol{\psi}(\tau_{ij}) \right\|_2^2 \tag{10}$$

where $\text{vec}(\mathbf{S}) \in \mathbb{C}^{ML\times 1}$ is the column-wise vectorized $\mathbf{S}$, $\otimes$ denotes the Kronecker product and

$$\mathbf{a}(v_i) \otimes \mathbf{M} \sum_{j=1}^{K_{v,i}} \alpha_{ij}\boldsymbol{\psi}(\tau_{ij}) = \mathbf{a}(v_i) \otimes \mathbf{M}\boldsymbol{\psi}_i = \text{vec}\left(\mathbf{M}\boldsymbol{\psi}_i \mathbf{a}(v_i)^T\right)$$

The problem (10) is a special nonlinear least square with the unknown $K_{v,i}$, and can be resolved by the compressive parameter estimation techniques [3, 9-11].

Directly computing (10) is time-consuming because of its high observation dimension $ML$ in $\text{vec}(\mathbf{S}) \in \mathbb{C}^{ML\times 1}$. We find that the problem (10) can be further decomposed into $K_v$ sub-problems with each corresponding to a Doppler shift $v_i$ and having low observation dimension $M$.

Note that $\boldsymbol{\Theta}$ is a Vandermonde matrix having full-row rank. Then $\boldsymbol{\Theta}\boldsymbol{\Theta}^\dagger = \mathbf{I}$ with $\boldsymbol{\Theta}^\dagger = \boldsymbol{\Theta}^H (\boldsymbol{\Theta}\boldsymbol{\Theta}^H)^{-1}$, the Moore-Penrose pseudo-inverse of $\boldsymbol{\Theta}$. Post-multiplying the both sides of the equation (7) by $\boldsymbol{\Theta}^\dagger$, we have

$$\mathbf{S}\boldsymbol{\Theta}^\dagger = \mathbf{M}\boldsymbol{\Psi} + \mathbf{N}\boldsymbol{\Theta}^\dagger = \left[\mathbf{M}\boldsymbol{\psi}_1, \mathbf{M}\boldsymbol{\psi}_2, \cdots, \mathbf{M}\boldsymbol{\psi}_{K_v}\right] + \mathbf{N}\boldsymbol{\Theta}^\dagger \tag{11}$$

It is seen that the $i$-th column of $\mathbf{S}\boldsymbol{\Theta}^\dagger$ is the noisy version of the measurement vector $\mathbf{M}\boldsymbol{\psi}_i$ corresponding to the Doppler shift $v_i$. Denoting $\mathbf{s}_{v_i}$ and $\mathbf{n}_{v_i}$ respectively as the $i$-th column of $\mathbf{S}\boldsymbol{\Theta}^\dagger$ and $\mathbf{N}\boldsymbol{\Theta}^\dagger$, we have

$$\mathbf{s}_{v_i} = \mathbf{M}\boldsymbol{\psi}_i + \mathbf{n}_{v_i}, \quad i = 1, 2, \cdots, K_v \tag{12}$$

Then directly solving (10) is equivalent to solving the $K_v$ sub-problems in which each is confined to find the delays $\tau_{ij}$ and gains $\alpha_{ij}$ corresponding to the Doppler $v_i$,

$$\arg\min_{\alpha_{ij},\tau_{ij}} \left\| \mathbf{s}_{v_i} - \mathbf{M} \sum_{j=1}^{K_{v,i}} \alpha_{ij} \boldsymbol{\psi}(\tau_{ij}) \right\|_2^2, \quad i = 1, 2, \cdots, K_v \tag{13}$$

In comparison with (10), the observation dimension of (13) is reduced $L$ times and thus the computational load is greatly alleviated.

In the CoSaPD method [4], a vector similar to $\mathbf{s}_{v_i}$ is generated by the row-wise DFT of $\mathbf{S}$. For the targets whose Doppler shifts are on the Nyquist grids, the matrix $\boldsymbol{\Theta}$ is a partial Fourier matrix and the DFT operation simply classifies the targets into groups with each corresponding to a distinct Doppler shift. For the formulation (7), the difficulty is that the matrix $\boldsymbol{\Theta}$ is not a partial Fourier matrix. However, noting the Vandermonde structure of $\boldsymbol{\Theta}$, we devise a similar decomposition as described in (11). This decomposition greatly enhances computational efficiency.

In addition to its generality to deal with the targets of the on- and/or off-grids, the vector $\mathbf{s}_{v_i}$ generated by (12) has the improved signal-to-noise ratio (SNR) with respect to the signal vector in (5). Let $\mathbf{b}(v_i)$ be the $i$-th column of $\boldsymbol{\Theta}^\dagger$. Then for each distinct Doppler shift $v_i$, since $\boldsymbol{\Theta}\boldsymbol{\Theta}^\dagger = \mathbf{I}$, we have

$$\begin{cases} \mathbf{a}(v_i)^T \mathbf{b}(v_i) = 1 \\ \mathbf{a}(v_{i'})^T \mathbf{b}(v_i) = 0, \quad i' \neq i \end{cases} \tag{14}$$

Let $\mathbf{P}_i = \mathbf{I}_N - \boldsymbol{\Theta}_i (\boldsymbol{\Theta}_i^H \boldsymbol{\Theta}_i)^{-1} \boldsymbol{\Theta}_i^H$ which is an orthogonal projection matrix with $\boldsymbol{\Theta}_i = \left[\mathbf{a}(v_1), \cdots, \mathbf{a}(v_{i-1}), \mathbf{a}(v_{i+1}), \cdots, \mathbf{a}(v_{K_v})\right]$ $\in \mathbb{C}^{L \times (K_v - 1)}$. With (14), we can derive

$$\mathbf{b}(v_i) = \frac{\mathbf{P}_i^* \mathbf{a}(v_i)^*}{\mathbf{a}(v_i)^T \mathbf{P}_i^* \mathbf{a}(v_i)^*} \tag{15}$$

By the definition of $\mathbf{n}_{\nu_i}$, we have

$$\mathbf{n}_{\nu_i} = \frac{\mathbf{N}\mathbf{P}_i^* \mathbf{a}(\nu_i)^*}{\mathbf{a}(\nu_i)^T \mathbf{P}_i^* \mathbf{a}(\nu_i)^*} \tag{16}$$

whose power is given by $\|\mathbf{n}'_{cs}\|_2^2 \|\mathbf{P}_i \mathbf{a}(\nu_i)\|_2^2 / |\mathbf{a}(\nu_i)^T \mathbf{P}_i^* \mathbf{a}(\nu_i)^*|^2$. Note that the signal power in (5) is the same as in (12) for a distinct Doppler target. Then after the decomposed processing, the SNR in (12) will be improved by $|\mathbf{a}(\nu_i)^T \mathbf{P}_i^* \mathbf{a}(\nu_i)^*|^2 / \|\mathbf{P}_i \mathbf{a}(\nu_i)\|_2^2 \leq L$. When all the Doppler shifts are on the Nyquist grids, all $\mathbf{a}(\nu_i)$ ($i=1,2,\cdots,K_\nu$) are orthogonal to each other and $\mathbf{P}_i$ is an identity matrix. Then $|\mathbf{a}(\nu_i)^T \mathbf{P}_i^* \mathbf{a}(\nu_i)^*|^2 / \|\mathbf{P}_i \mathbf{a}(\nu_i)\|_2^2 = L$ and the SNR for the on-grid targets is enhanced by $L$ times, which is consistent with the analyses on CoSaPD in [4]. For CoSaPD, the row-wise DFT operation performs as matched filtering processing by $\mathbf{a}(\nu_i)^*$ in Doppler dimension. In this sense, the decomposed processing to generate the signal vector $\mathbf{s}_{\nu_i}$ has the similar function. However, for the off-grid targets, an extra pre-processing $\mathbf{P}_i^* / (\mathbf{a}(\nu_i)^T \mathbf{P}_i^* \mathbf{a}(\nu_i)^*)$ is introduced, which cause a loss on the SNR improvement.

In classic pulse-Doppler processing, we usually perform the matched filtering in time domain and coherent integration in Doppler dimension to enhance the target strength for further processing. One may argue if there is SNR loss for the proposed scheme because the matched filtering processing in time domain does not explicitly appear, as pointed out by one of reviewers. As discussed in other works [4], this is a difference of the radar signal processing in compressed domain from the classical processing. In fact, the matched filtering processing is implied in the estimation process by (13), which increases the input SNR by $MBT_p/N$ times. Then the system processing gain after the delay estimation will be improved to be $\left[ |\mathbf{a}(\nu_i)^T \mathbf{P}_i^* \mathbf{a}(\nu_i)^*|^2 / \|\mathbf{P}_i \mathbf{a}(\nu_i)\|_2^2 \right] (MBT_p/N)$. For the on-grid targets, the processing gain equals to $LMBT_p/N$, which is same as that in CoSaPD.

The proposed scheme performs the estimation of the Doppler shifts and the delays sequentially by (8) and (13). Although the SNR is greatly enhanced at each processing stage, the system processing gain by the proposed sequential processing is less than that by the classic pulse-Doppler processing because of the noise-folding phenomena caused by the sub-Nyquist sampling. Then the optimal estimation accuracy with the compressed data in (7) is lower than that by the Nyquist data as revealed by the Cramer-Rao bound (CRB) analyses in [20]. The simulations in Section V confirms

## IV. ANALYSIS OF COMPUTATIONAL COMPLEXITY

This section analyzes the computational complexity of the proposed scheme. We utilize ESPRIT algorithm [17] to estimate the Doppler shifts because of its high estimation accuracy and computational efficiency, and exploit the parameter perturbed orthogonal matching pursuit (PPOMP) algorithm [3] to estimate the delays and gains for its high estimation accuracy.

As is known, estimating the Doppler shifts from $\mathbf{S}$ by ESPRIT algorithm takes $\mathcal{O}(ML^2 + L^3)$ operations. To estimate the delays and gains, we take $\mathcal{O}(KLM + K^2L + K^3)$ operations to generate all the vectors $\mathbf{s}_{v_i}$ from $\mathbf{S}$ and $\mathcal{O}(PMN)$ operations to solve the problem (13) for each $\mathbf{s}_{v_i}$ where $P$ is the maximum number of iterations for nonlinear optimization sub-problems in PPOMP [3]. In the worst case where the delays and Doppler shifts are distinct between any two targets, we have to solve the problem (13) by $K$ times. Then we have the total computational load of the proposed scheme as $\mathcal{O}(ML^2 + L^3) + \mathcal{O}(KLM + K^2L + K^3) + \mathcal{O}(KPMN)$. For sub-Nyquist radar, it is known that $K < L < P \ll M < N$. Then the computational complexity of our scheme is approximated as $\mathcal{O}(KPMN)$. In [7], we have derived that the complexities of GeSeDD via beamspace root MUSIC (abbreviated as GeSeDD-1) and GeSeDD via beamspace spectral MUSIC (abbreviated as GeSeDD-2) are respectively approximated as $\mathcal{O}(N^3)$ and $\mathcal{O}(DMN^2)$, where $D$ is the ratio between the Nyquist grid space and the grid-search space and generally satisfies $K < D < L$. It is also argued in [7] that the computational complexity of GeSeDD is comparable to that of CoSaPD method and Doppler focus (DF) method in [5]. Then the proposed scheme has the smallest computational complexity among these methods.

## V. SIMULATION RESULTS

This section simulates some experiments to show the estimation accuracy and the computational advantages of the proposed scheme. The transmitted signal is assumed to be a linear frequency modulation pulsed signal with bandwidth $B = 100\,\text{MHz}$ and pulse width $T_p = 10\,\mu\text{s}$. The PRI is $T = 100\,\mu\text{s}$ and the number of pulses is $L = 100$. The delay cell size $\tau_0 = 1/B$ is $0.01\,\mu\text{s}$ and the Doppler cell size $v_0 = 1/(LT)$ is $0.1\,\text{KHz}$. Without special statements, the delay and the Doppler shift of all the targets are randomly chosen with a uniform distribution from the intervals $(0,10)\,\mu\text{s}$ and $(-5,5)\,\text{KHz}$. The gain amplitudes and phases of the targets are uniformly distributed in $[0.1,1]$ and $(0,2\pi]$, respectively. The $\text{SNR} = \|\mathbf{s}_{cs}^l - \mathbf{n}_{cs}^l\|_2^2 / \|\mathbf{n}_{cs}^l\|_2^2$ is assumed to be identical in each pulse interval. The estimated delays and Doppler shifts are determined by the estimated targets having the $K$ largest amplitudes of gains.

The Xampling is taken as a sample AIC with one-fifth Nyquist rate, where the measured Fourier coefficients are randomly selected. For performance comparisons, simulation results by GeSeDD-1 and GeSeDD-2 [7], CoSaPD [4] and DF [5] are demonstrated. As a benchmark, a simultaneous method, the PPOMP in [3], is also shown for its high estimation accuracy. For GeSeDD-2, the grid space is set as one-tenth delay cell size for grid search in beamspace spectral MUSIC. For the DF method, the grid spaces of delay and Doppler shift are both set as a half of cell size.

We now evaluate the delay/Doppler estimation performance of all methods in the noisy case. The relative root-mean-square error (RRMSE) is utilized as a metric for the performance evaluation. Let the set $\bar{\boldsymbol{\tau}} = \{\bar{\tau}_1, \bar{\tau}_2, \cdots, \bar{\tau}_K\}$ and $\bar{\mathbf{v}} = \{\bar{v}_1, \bar{v}_2, \cdots, \bar{v}_K\}$ be the delay and Doppler shift estimate values of $K$ targets. The RRMSEs of the time delay and the Doppler shift are defined respectively as

$$\text{RRMSE}_\tau = \frac{1}{\tau_0}\sqrt{\frac{1}{K}\sum_{k=1}^{K}(\bar{\tau}_k - \tau_k)^2} \quad \text{and} \quad \text{RRMSE}_v = \frac{1}{v_0}\sqrt{\frac{1}{K}\sum_{k=1}^{K}(\bar{v}_k - v_k)^2}$$

In the simulation, the delay and Doppler separation between any two targets are respectively set to be at least $2\tau_0$ and $2v_0$ to alleviate interaction between different targets. Fig. 1 displays the curves of the RRMSEs versus the SNRs with ten off-grid targets for the noisy compressed data. The CRBs of delays and Doppler shifts are provided as a benchmark, which can be derived from the vectorized representation of (7). It is seen that when $\text{SNR} \geq -10\,\text{dB}$, the proposed scheme has the same accuracy as GeSeDD-1 and is superior to the other sequential methods. However, both the proposed scheme and GeSeDD have a sharp increase in RRMSE when $\text{SNR} < -10\,\text{dB}$. This is because both schemes exploit subspace DOA estimation techniques to find the Doppler shifts or the delays. As we all know, the subspace techniques suffer a threshold effect [7], i.e., the estimation performance is sharply decreased when the SNR is below a threshold value. For a small SNR ($\text{SNR} \leq -10\,\text{dB}$), PPOMP offers the highest estimation accuracy because of its robustness to noise [7]. In spite of the SNR, both CoSaPD and DF have poor estimation accuracy due to the basis mismatch. As analyzed in [7], GeSeDD-1 has the best estimation accuracy among the simulated sequential methods. This simulation shows that the proposed scheme has high estimation accuracy and is comparable to any other methods.

The proposed scheme is also applicable to the Nyquist data for the estimation of the off-grid targets. We simulate the RRMSE performance of the classic pule-Doppler processing [12]. A quadratic interpolation technique [12] is exploited to improve the estimation accuracy of the delay and Doppler shift parameters, where the Nyquist data is windowed by a hamming window. Fig.2 shows the estimation performance along with CRBs from the Nyquist data. We see that the proposed scheme has the best estimation accuracy which approaches to the CRBs for the higher SNRs. In the classic processing, the interpolation techniques can improve the estimation accuracy. However the improvement is limited as discussed in [12]. The CoSaPD method has the worst performance because it works only for the on-grid targets. This further shows the efficiency of the proposed parametric scheme. In comparison with Fig.1, we find that the classic processing performs better among all the simulated schemes with sub-Nyquist data at low SNRs ($\text{SNR} \leq -15 \text{ dB}$) because of its higher processing gain. Moreover, we can find that the CRB from the Nyquist data is lower than that from the compressed data, which is consistent with the analyses in [20]. For the proposed scheme, the estimation accuracy from the compressed data is also lower than that from the Nyquist data. Then the pulse-Doppler processing in the compressed domain reduces the estimation accuracy to a certain extent. However, the reduction is negligible, as shown in Fig. 1 and Fig. 2.

Another advantage of the proposed scheme is that it takes less computational time. We take the CPU time to illustrate the performance. The simulation is performed in MATLAB 2011b 64-bit environment on a PC with 3.6 GHz Intel core i7-4790 processor and 16 GB RAM. Fig. 3 displays the dependence of the CPU time on the number of targets $K$ with SNR=20dB. It is clear that the proposed scheme has the least computational cost. This further verifies our analyses in Section IV. Note that the CPU time of the proposed method increases along with larger $K$. This is because the computational complexity is dominated by the delays and gains estimation and is linearly proportional to $K$, as analyzed in Section IV. Similarly, the DF and CoSaPD methods lie on $K$, and the DF has stronger dependence on $K$. On the contrary, the CPU time of both GeSeDD methods is independent on $K$. The reason is that the computational cost is controlled by the beamspace DOA estimation problem, which is not affected by $K$ [7].

## VI. CONCLUSION

In this paper, we have presented a new general sequential estimation scheme for estimating the sub-Nyquist radar parameters. Different from our previous contributions [7], we first estimate the Doppler shifts and then the delays and gains of the targets. With the new mechanism, the proposed scheme has the less computational complexity and the high estimation accuracy among all methods reported. Simulations validate our theoretical development. This sequential estimation idea has been used in the development of the CoSaPD method [4]. The difference is that, unlike [4] in which the targets are assumed to be on the predefined grids, the new scheme developed in this paper removes this constraint. In this sense, the proposed scheme is a gridless CoSaPD.

We are currently working on the development of the target detection techniques incorporated into the proposed sequential scheme.

## ACKNOWLEDGEMENTS

The authors would like to thank the anonymous reviewers whose constructive comments improved the content and the presentation of this paper. This research is supported partially by the National Natural Science Foundation of China (Nos. 61401210, 61571228 and 61671245).

## REFERENCES

[1] M. A. Herman and T. Strohmer, High-resolution radar via compressed sensing, IEEE Trans. Signal Process. 57 (6) (2009) 2275-2284.
[2] J. Zhang, D. Zhu, and G. Zhang, Adaptive compressed sensing radar oriented toward cognitive detection in dynamic sparse target scene, IEEE Trans. Signal Process. 60 (4) (2012) 1718-1729.
[3] O. Teke, A. C. Gurbuz, and O. Arikan, A robust compressive sensing based technique for reconstruction of sparse radar scenes, Digital Signal Process. 27 (4) (2014) 23-32.
[4] C. Liu, F. Xi, S. Chen, Y. D. Zhang, and Z. Liu, Pulse-doppler signal processing with quadrature compressive sampling, IEEE Trans. Aerospace and Electronic Systems 51 (2) (2015) 1216-1230.
[5] O. Bar-Ilan and Y. C. Eldar, Sub-Nyquist radar via Doppler focusing, IEEE Trans. Signal Process. 62 (7) (2014) 1796-1811.
[6] W. U. Bajwa, K. Gedalyahu, and Y. C. Eldar, Identification of parametric underspread linear systems and super-resolution radar, IEEE Trans. Signal Process. 59 (6) (2011) 2548-2561.
[7] S. Chen, F. Xi, and Z. Liu, A general sequential delay-Doppler estimation scheme for sub-Nyquist pulse-Doppler radar, Signal Process. 135 (2017) 210-217.
[8] P. Stoica and R. L. Moses, Spectral Analysis of Signals. Upper Saddle River, NJ, USA: Prentice-Hall, 2005.

[9] C. Ekanadham, D. Tranchina, and E. P. Simoncelli, Recovery of sparse translation-invariant signals with continuous basis pursuit, IEEE Trans. Signal Process. 59 (10) (2011) 4735-4744.

[10] K. Fyhn, M. F. Duarte, and S. H. Jensen, Compressive parameter estimation for sparse translation-invariant signals using polar interpolation, IEEE Trans. Signal Process. 63 (4) (2015) 870-881.

[11] A. Fannjiang and W. Liao, Coherence pattern-guided compressive sensing with unresolved grids, SIAM J. Imaging Sci. 5 (1) (2012) 179-202.

[12] M. A. Richards, Fundamentals of Radar Signal Processing. New York, NY, USA: McGraw-Hill, 2005.

[13] F. Xi, S. Chen, and Z. Liu, Quadrature compressive sampling for radar signals, IEEE Trans. Signal Process. 62 (11) (2014) 2787-2802.

[14] R. G. Baraniuk and P. Steeghs, Compressive radar imaging, in Proc. IEEE Radar Conf., 2007, pp. 128-133.

[15] F. Xi, S. Y. Chen, and Z. Liu, Quardrature compressive sampling for radar signals: Output noise and robust reconstruction, 2014 IEEE China Summit and International Conference on Signal and Information Processing, 2014, pp. 790-794.

[16] R. O. Schmidt, Multiple emitter location and signal parameter estimation, IEEE Trans. on Antennas Propagat. 34 (3) (1986) 276-280.

[17] R. Roy and T. Kailath, ESPRIT-estimation of signal parameters via rotational invariance techniques, IEEE Trans. on Acoust., Speech, Signal Process. 37 (7) (1989) 984-995.

[18] S. U. Pillai and B. H. Kwon, Forward/backward spatial smoothing techniques for coherent signal identification, IEEE Trans. on Acoust., Speech, Signal Process. 37 (1) (1989) 8-15.

[19] F. Xi, S. Chen, Y. D. Zhang and Z. Liu, Gridless quadrature compressive sampling with interpolated array technique, Signal Process. 133 (2017) 1-12.

[20] P. Pakrooh, A. Pezeshki, L. L. Scharf, D. Cochran, and S. D. Howard, Analysis of Fisher information and the Cramer-Rao bound for nonlinear parameter estimation after compressed sensing, IEEE Trans. Signal Process. 63 (23) (2015) 6423-6428.

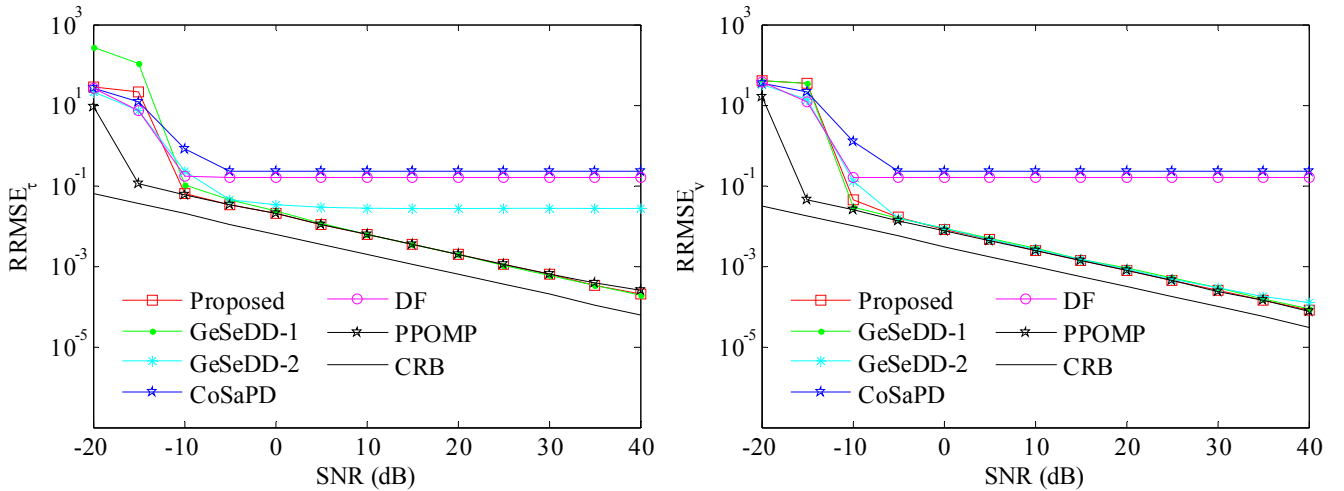

(a) time-delay       (b) Doppler shift

Fig. 1 The RRMSEs versus the SNR for noisy compressed data ($K$=10)

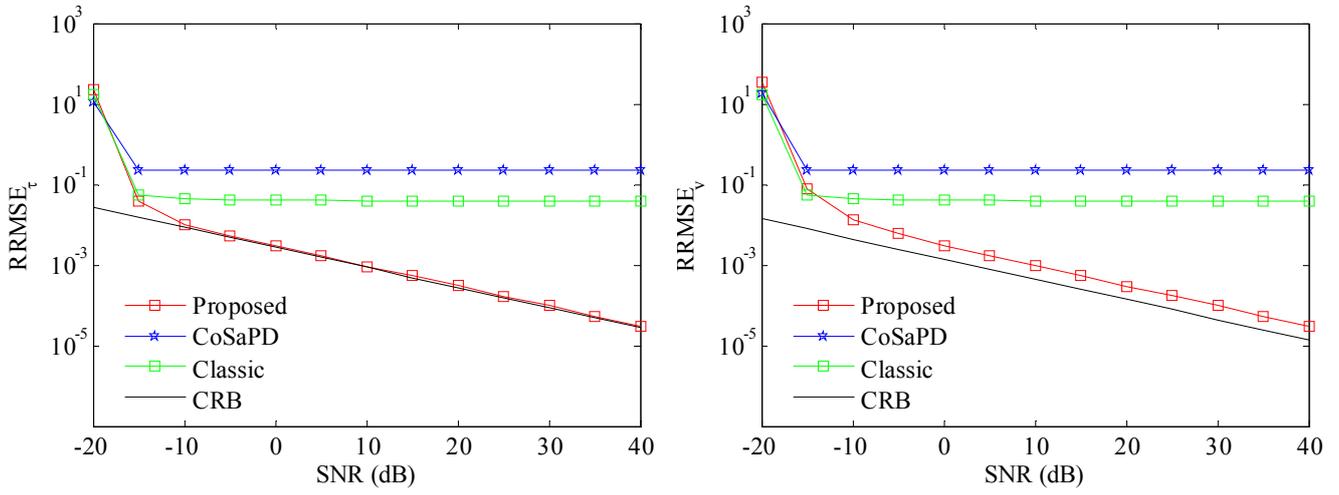

(a) time-delay    (b) Doppler shift

Fig. 2 The RRMSEs versus the SNR for noisy Nyquist data ($K$=10)

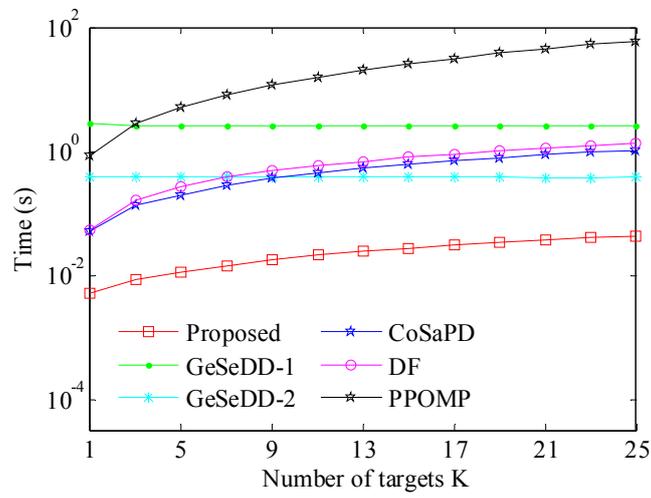

Fig. 3 The CPU time versus the number of targets with SNR=20dB